# Domain Specific Design Patterns: Designing For Conversational User Interfaces

Ahmed Fadhil
*University of Trento*
*Trento, Italy*
fadhil@fbk.eu

*Abstract*—Designing conversational user interface experience is complicated because conversation comes with many expectations. When these expectations are met, we feel the interface is natural, but once violated, we feel something is amiss. The last decade witnessed human language technologies and behaviours to enable humans converse with software using spoken dialogue to access, create and process information. Less is known about the practicalities of designing chatbot interactions. In this paper, we introduce the nature of conversational user interfaces (CUIs) and describe the underlying technologies they are based on. Moreover, we define guidelines for designing conversational interfaces in various domains. This paper particularly focuses on classifying the elements and techniques used in CUI design patterns. After concluding certain challenges with CUI, we discuss important features and chatbot states to be considered in CUI design for specific domain. We envisage this study to support CUI researchers to design tailored chatbots applicable into certain domain and improve the current state of research challenges in the field of Artificial Intelligence and conversational agents.

*Keywords*-User Experience, User-centred Design, Design Patterns, Chatbots, Conversational UI, AI

## I. INTRODUCTION

Conversational User Interface (CUI) is a software that runs simple and structurally repetitive tasks inside a messaging application [1]. Conversation is how we share knowledge, emotions and it has been part of our makeup for years. Moreover, messaging applications are used in everyday life and are becoming a layer in our daily life. Chatbots are expected to be the new generation of digital product design, after the evolution from Web to mobile applications. The reason lies behind the simplicity offered by the CUI when performing a task, which otherwise could be tedious or requires more time and effort if performed via web or mobile apps. In addition, users spend more time on their messaging platforms than any mobile apps, this generates a new shift to address user needs. Introducing GUI to the web and mobile makes it less efficient and tedious to perform a task [2]. To illustrate, it takes around 18 clicks to perform an airline reservation, while we're faced with an unwieldy array of buttons, ads, drop-downs, text boxes and more. Simplicity is crucial when communicating with our device. Rather than pulling up an app to search for restaurant, tap to select time and type in number of people, we can simply tell a chatbot to "Book me a table for three at 6 tonight, at the Elzar's place". We could facilitate a conversation between a person and a service to fulfill users' request [2]. Using a chatbot to access a service wont requires users to familiarise themselves with the bot as it is the case with mobile apps. Moreover, developing a bot requires less time and cost to build and deploy than a web or mobile app. Consequently, currently the most popular apps in the world are either social in nature or primary for messaging. Although this new paradigm shift, the current state of conversational interface is limited in terms of established user interface design patterns. The current resources on CUI design and the specific bot behaviour in certain domain is limited. In addition, it's unclear when bots should be text or button and keyboard based. However, the big question remains how to use this new medium to build great user experience. It is important to consider users' point of view, such as their needs, what motivates them to seek your bot and how to create unique CUIs for various users and domains when designing CUIs. We witness a transformation represented by chatbots for messaging apps, similar to what API's were for Web 2.0. This is a new way to build services and create interaction experiences for users. This paper investigates application domains for conversational interfaces and defines techniques and features per domain and describes a way to implement domain specific CUI features. We particularly cover user needs and interaction with the chatbot in a specific domain. We contribute to the definition of domain specific CUI design patterns from the UX point and the existing knowledge on CUI design. Moreover, it will set the roadmap to define best practices and address CUI design challenges for researchers and developers to follow when designing domain specific CUIs. We will investigate the role of design in CUI and what it means to design for chatbot experience.

## II. CONVERSATIONAL INTERFACES

Although CUIs are rather small sight compared to web and mobile apps, more applications are switching from GUI in favour of conversational interface. The reason is messaging is growing as a platform, bots are easy to install and can deliver unique user experience. Effective bots are context aware and can retrieve past interaction and use the data in the current context. Current CUIs are driven either by artificial intelligence or have a human facilitating some tasks behind the scene. This is to enhance user experience since it feels that the bot is starting to understand them. Bots living on messaging apps don't need the user to download anything. CUIs are based on text dialogue and buttons, although some platforms have added new services and features to their existing one. For example, the gaming platform offered by Telegram bot API. To create effective interaction design, a balance between the text dialogue and the custom keyboard is necessary as a source of input and part of the CUI [2]. Bots personality is important to understand how it can build empathy and emotional connection with the user (See Table-I for a comparison of the major differences between the native vs bot applications).

Table I: Major Advantages Of Conversational Interfaces Over Mobile Applications.

| Mobile Applications | Chatbot Applications |
|---|---|
| Accessed only when required | Used daily to send/receive messages with other users |
| Requires more time and resources | Saves time and cheaper to develop |
| GUIs require learning | CUIs are intuitive and simple to grasp |
| Requires updates | No app update required |
| Requires high-connectivity bandwidth | Messages are suitable for low-connectivity situations |
| Requires an app per service | Single bot to roll them all |
| Requires time to get used to it | Less friction, easy to use from the first time |
| Requires creating account, downloading app, and learning new interface | No account, no download and no interface learning required |
| Provides restricted functionalities | Allows more flexibility |

### A. Scenario Description

To better present the major differences between a mobile and chatbot application, we apply it to an example of providing a service to the user, first with a mobile application, then by using conversational interface.

**Mobile Application:** *John would enjoy having a popcorn while enjoying the match in the stadium. He notices the stadium has announced about popcorn ordering online. John unlocks his phone, goes to google play, searches for an app, puts his password, waits for it to download, creates an account, enters his credit card details, figures out where he actually orders from. Now he has to figure out where to place the order and quantity. Finally, he has to insert his seat number, only after that John can submit the popcorn order.*

**Chatbot Application:** *However, if we consider the same scenario, except that the stadium decides to spend up to a thousand to develop a simple, text-based bot. This time John sees a sticker "Want a popcorn?, chat with us" with a barcode beside it. Now John unlocks his phone, opens his chat app, scans to find the bot. Instant, he begins chatting with a stadium bot, and would place the order type and quantity. Finally, the bot would ask for the payment type, and Johns popcorn will be on its way.*

### III. GENERAL CUI DESIGN PRINCIPLES

It is important for product designers to intuitively understand what a first-time user may look like, and how each user's learning curve is unique. This design philosophy will optimise the product around how user can, want or need to use a product, rather than forcing them to change their behaviour to accommodate the product.

**Personality by design:** The chatbot personality is what gives the feeling of a natural conversation with an individual and comfort in the conversation. Designing the appropriate personality can increase the user experience and engagement with the bot. Personality is the tone the bot takes in a conversation. For example, if the bot is intended for e-commerce, then the bot personality should cover features, such as understanding the context of the conversation, is the user browsing or looking for specific information. Moreover, personality also includes knowing what the user might ask for and what the bot can't offer.

There is no one-size-fits-all approach when it comes to designing the correct conversation flow and currently there are no patterns that work for all cases. Most bots lack the structure to steer a conversation from the beginning to the end. Bots personality has to defines steps to guide user to learn and manage their intention. The bot should always be one step ahead of the user. For example, the bot should state which topics it covers when greeting the user, as below:

```
Bot: I can help you to book a table, select menu
and pay online.
```

This guiding is helpful and will result in one of the provided options being selected by the user. We need to understand what motivates us within a conversation and how can this be programmed into chatbots behaviour. Personality in chatbot is the new UX within conversational UIs [3]. Bot personality have to reflect the specific domain it is employing. For example, if we have a bot to take pizza orders, then one possible personality reflection could be that a significant number of people are asking for "Pizza

Pepperoni", which is not available. In this case, the bot has to provide clever responses that also progresses the conversation, instead of falling into a dead-end, as illustrated below.

```
Bot: You can select pizza pepperoni, pizza
margherita and pizza with veggies. What is your
choice?
Human: I want pizza pepperoni
Bot: 1 pizza pepperoni, cool, now tell me the size
(small, medium, large)?
```

```
Bot: Hi John, it's time you check your daily
dietary plan. Open the application for details.
```

**Flexibility in Response:** The bot should provide flexible values to various user requests. To illustrate, providing different error messages as response to the same question arose by the user. Moreover, the bot should cover irregular cases, such as a keyword related to another branch of the decision tree or a completely irrelevant keyword to the context. For instance, asking about a service before getting confirmation of success during a payment and causing early termination. If the user asks a random question or tricks the bot with unrelated questions, then it is critical that the bot doesn't repeat itself with a response such as:

```
Bot: Sorry I didn't quite get that!
```

If the bot continues to provide no information on alternate course of action, then the probability a user will leave the bot is very high. A best strategy is to play with multiple types of response or error messages and see what gets the best response from your user by focusing on a humorous or satirical approach to the problem. More importantly, provide different error messages for specific errors that occurs in a conversation. If the error occurs when the user is asked to provide some information, such as their address, then the bot could respond:

```
Bot: Sorry I'm still learning, try formatting your
address 231 Cedar St, NY 11211
```

Moreover, if a user encounters an error while asking for information the bot could respond:

```
Bot: Hey! Not sure we have that on the menu. You
can see everything we offer here: "link_address".
```

**Text vs Custom Buttons:** Users shouldn't be placed in a situation where they have to guess the correct information required to proceed. Moreover, the bot should support a dictionary databased for synonyms so to yield the same result for vocabularies, such as "buy" and "purchase" or "client" and "customer". Custom keyboards or buttons permit a limited range of inputs and can save a bunch of typing. To illustrate, rather than asking end user to type "Yes" or "No" text, the bot can present them with two mutually exclusive buttons. Moreover, the bot can validate structured text link email address before sending. This way allows the bot to keep responses on track and sidestep the complications of parsing unpredictable plain text input. They bot should never leave the user clueless in case of unrecognised questions and provide a fallback response to guide user as to what the next step might be. For example, provide the main topics from the content domain available. Rather than just using free-text NLP communication with the bot, a proper UI should present options to the user to guide them through a decision-making process. For instance, if the user is checking for a jacket, the interface can suggest a few of the company's latest styles and similar items, as given below.

```
Human: I want to buy a snow jacket
Bot: We have item1, item2, item3
```

**Simplicity in Interaction:** Beyond the seemingly human aspect of communicating in the right tone and predicting and learning the user through a task, an additional reason CUI will work well in domains, such as healthcare is their simplicity. Since this extremely removes the layers of abstraction that was introduced with interface design patterns. Which is often slow and more thoughtful, and which works in the right situations. However, when designing for emotional consideration, perhaps even looking to add little thoughtful friction back into the exchange so that the user feels like someone's listening then CUI is pretty great. Conversational interfaces has to be bounded to particular subjects and follow a linear conversation routes [4]. Users often are looking for a service that lends them a hand when they need information, find them the answer with no extra effort. Therefore, individual designers should avoid complicated branching paths and shouldn't have to account for tricky failure cases. Moreover, utilising auto-complete whenever possible and suggesting questions with the same wording are both effective to steer and inform users. This will shorten typing effort, since users can select one for the items in the auto-complete list.

```
Human: I need a taxi
Bot: I can offer you these options for a taxi:
opt1, opt2, opt3
```

**Conversation Flow:** Question and answer conversational interfaces are not the most effective way of getting answers from a bot, no matter how intelligent they are. For example, when ordering a food through a bot and when user isn't sure what to order, then they have to go through a conversation with the bot to figure out what to order. The conversation path becomes important, since it can help figure out how the conversation flow should be designed (e.g., tree hierarchy).

Moreover, if the user can't find answer to his question, it makes sense to add an option for the user to provide a feedback to save user question. The point is to prevent user from getting frustrated and provide a guidance instead of repeatedly saying "Sorry, I did not understand that", as below.

```
Human: how many toppings I can choose for my
pizza?
Bot: We offer the following pizza types: opt1,
opt2, if you have a request please insert it in
the below box
```

**Tasks and Duty Specifications:** Chatbots should serve specific and clearly designed tasks, since otherwise, bots that try to do too much usually fail. To illustrate, Viv and Siri bot suffer from aiming to cover everything, and in the process compromise on quality. To avoid this issue, we need to understand how chatbots work. Below is an illustration of bot for specific purpose.

```
Human: Hi
Bot: Hi John, I can help you book/cancel a doctor
appointment and communicate with your general
physician.
```

**Rigid Syntax, then NLP:** Starting with rigid syntax, then introducing NLP is a best practice to start with when designing CUI. The issue with NLP is that even if it supports complex conversation and gets the sentences right in most cases, it will fail in the remaining sentences in a very predictable way. Even if the bot provides fun error messages and smart responses to various user questions, this failing might lead to user frustration. NLP is perhaps the biggest bottleneck when designing CUIs. Humans always do spelling, grammar or typo mistakes. In addition, users often use slang and nuance to express something which can make any NLP based system to struggle. In fact, some services rely on employees to manage their conversation UI. NLP is effective in specific tasks, it is tedious to type and error prone. Therefore, focusing on creating a good UI is more important than a complex NLP. There are a number of ways in which NLP falls short of the requirements for a good UX by comparison with traditional UIs. For one thing, people are lazy and it feels easier to click a button, as we do on apps and websites, than typing out the whole sentence that the bot may will not understand. In addition, people easily fall into typo mistakes and the text they input may be full of grammar and spelling mistakes and colloquialisms that are bound to confuse the AI system. This will result in user frustration with the bot in the long-term, since it can't understand their demands. Even worse, it may do the opposite of what it was asked for, as such:

```
Human: Please stop sending me offers
Bot: Here are some new offers
```

Where the NLP reads "Please stop sending me offers" and interprets it as "Send me offers"; it then sends the user more offers.

**Empathy & Emotional State:** One of the pillar in CUI design patterns is to consider building an empathy with the user. All the design should be approached with empathy, in fact, many CUIs integrate pauses, humans take a second to type an answer, so it makes sense to build a typing indicator for a bot. Bots work really well when there is a decision tree. Many developers and platforms are working on bots emotional sensitivity needed to make complicated judgements that can't be captured without a decision tree. Addressing a social issue requires emotional sensitivity, a critical skill that bots are universally missing. Bots have to present some aspects of emotional state changes (fear, anger, sadness, etc.). In this regard, LawBot [5] which is a legal bot created by Cambridge University students to help users understand the complexities of the law and identify whether a crime has been committed. The bot is used to report rape, sexual harassment, injuries and assaults, and property disputes. However, the bot relies on strict role based checklist to assess if a crime has been committed. If the user reports sexual harassment, and it doesn't fit within present criteria, the bot response with the following:

```
Bot: I don't think a sex offence was committed
here. Say "Crime" for a list of what I can help
you with.
```

Despite good intentions, the emotionally insensitive LawBot quickly dismisses sexual harassment if the harassment does not fit within a narrow set of legal technicalities. The results can be counterproductive and further discourage victims to speak up.

**Keep Conversation Short:** Human-bot interaction should be short and precise. Introducing a protracted back and forth conversation will make it feel laborious and hard to interact with the bot. Instead, bots have to serve specific tasks and intend for specific domains, in other words, personalised for a specific purposes. Unlike, GUI that defines rules for each interaction which often frustrate users, CUI has to be liberating in their familiarity.

```
Bot: We offer S, M, and L T-shirts, and currently
have only black and green colour available. Please
select from the button.
Human: Selected (S-Black)
```

**Triggers and Actions:** It is crucial to work on the persuasive aspect of the bot and find a way to trigger the user to perform an activity via the bot application (e.g., motivate user to perform a purchase). One scenario is the patient example, when there is a steep drop-off on drug adherence for patients in the first 90 days. There are several associated reasons, it's hard to integrate something new into your daily routine or new side effect can be hard to manage. However, to persuade patients towards continuous medication, we need to figure out their needs and fears, then design to fit the right goals. We need to learn what patients value and try to connect to it with compassion and right moment [6].

```
Human: I feel depressed
Bot: Would you like to tell me about it, or talk
to Dr. James?
```

**Fully/Partially Automated:** Before introducing any functionalities into the bot, one should ask themselves whether a human will be better for the end user or a bot will deliver a more effective service. Bots are not to replace what humans are good at, rather they should improve what humans are slow at. As much advanced in ML and AI are going to help with automation where it makes sense, it is also true that we will absolutely need a human being in the loop to create the right end-to-end customer experience. It is a good practice to start with real humans, doing 100% of the work, then, introduce role-based hierarchical tree. Only after that, introducing NLP to automate the 10% of the most common tasks. As an example, introducing NLP to the most frequently asked questions or just for the on-boarding. At later stages, especially when we have enough dialogue data, then we can move up and introduce more automation and intelligent machine learning techniques. This strategy applies when the value of the bot is in the human expert.

```
Human: I need to book a flight
Bot: Hi John, I am Sam, your assistance. Where do
you want to fly?
```

However, when the value is in task automation, then start with a rigid syntax. This provides helpful reaction messages to guide user when they enter an incorrect format. Starting from rigid syntaxes makes it easier to introduce NLP to provide more flexibility in the sentence when accepting or handling errors. Integrating some humanised touches into automated replies will make users feel more comfortable engaging with the bot. On top of that, sometimes human expert is needed in the loop, especially in medical bots, to create the right end-to-end customer experience.

```
Human: I need to book a flight
Bot: Hi John, please choose your departure and
arrival points and dates from the keyboard.
```

**Predict & Personalize:** Asking information is a one way street, but with bots this will change. Bots can be programmed to ask specific and time sensitive information and get as many variables as possible from user. Bots can predict what the user's preferences might be, based on previous history, and hence provide the user with personalised information. To illustrate, bots in medical domain will be able to ask the same questions a doctor would, get personal information (e.g., symptoms) and provide user with accurate diagnoses. Further, bots can remember the preferences, medical history and other relevant data about user. This will increase their accuracy and speed in future sessions. Similarly, a fashion bot could take a number of variables and suggest the perfect outfit for the user. The bot can check for parameters, such as what is temperature like, your personal preferences, and even consider your mood. In conclusion, they can give us the right information just before we need it.

```
Human: I had high blood pressure today
Bot: Did you consider the dietary plan I gave you
regarding your diabetes restriction?
```

**Provide a Way Out:** Humans make mistakes, therefore, bot users should always be able to start over, make changes, or completely escape when a mistake occurs. Sometimes, humans get into a conversation without a correct established base facts. However, when it becomes clear that the other party is lost, we start over or stop the conversation, bots should do the same.

```
Bot: Would you like to order pizza?
Human: I am not sure what to eat tonight
Bot: How about a list of available restaurants to
order from?
```

**User Boredom:** Having considered user engagement in the conversation, it is reasonable to look at what causes this engagement to drop to result in user boredom. The best CUIs are the ones that offer a new experience every time a user comes back. This could be through the conversation and a progressively simplified UI or a streamlined checkout process but on previous ordering preferences. Hence, bot functionalities should give a way to delight the experience and transform a digital product from one that's tolerated into one that's friendly.

```
Human: Feeling bored
Bot: Would you like some funny gifs? just type
something you would like to see
```

## IV. DOMAIN SPECIFIC CUI DESIGN PATTERNS

Designing habits requires addressing specific domains with functionalities and features to fulfill certain action. Therefore, to design domain specific CUI, we have to address domain specific design pattern challenges, much like a pattern library for GUI. This section discusses the context where each design patterns might occur. We particularly discuss best practices in CUI design patterns for input validation, question/answer, and so on. We discuss the context, direct message and where a certain pattern fits. Below we list and discuss each of the pattern features useful when designing interfaces and defining CUI behaviours (See Table-II for the domain specific application of CUIs). Next, we list bot application domain by referencing designs for frequently used elements and techniques in CUI design.

GOAL-FULFILMENT: This feature is to create engaging bots that maintain user interaction. The interaction with the bot has to be natural and as human-like as possible. Therefore, to add goal-fulfilment feature into the bot, predict and personalise, analytics, flexibility and reaction to interaction elements have to be considered. Since these elements are responsible for the goal-fulfilment behaviour in a bot.

INPUT VALIDATION: To have an idea of user input, we need to know what is the conversation nature. Therefore, to validate user input we must consider the conversation medium (e.g., text, button, or graphics). Moreover, which technique the bot uses to process the message (e.g., role based, NLP, or some machine learning based). Therefore, we have to consider using text vs custom keyboard, and rigid or NLP syntax when working with CUI input validation.

ON-BOARDING: This is essential, especially for newcomers. The conversational interface has to provide a value for users with flexible and different responds for each user demand. In addition, it has to suggest commands, brief configuration steps and pass quickly to value-delivery step. To enrich these features in a bot, it should be provided with a conversation flow and emotional state elements. It should also provide flexible responds as a backup in case unrelated message is received and predict and personalise. These elements will help make a more natural conversation and provide a path that doesn't leave user with unanswered questions.

CONVERSATION FLOW: To provide a natural conversation flow, the conversation should be bounded to particular subject and follow a linear path. In addition, the interaction should be free of abstract layer and doesn't have to be limited to chat, action buttons and images have to be there too, since this shortens the path to perform an action (e.g., purchase completion). Introducing a protracted back and forth conversation makes it feel laborious and hard to interact with bot. Chatbots have to serve specific tasks and intend for specific domain. Techniques, such as simplicity, provide a way out, conversation flow, bot flexibility, short conversation and providing reaction to interaction can provide a robust conversation.

FOLLOW-UP QUESTIONS: Bots have to ask a set of domain specific questions. These questions depend on situational context in the domain. The number of follow-up questions have to be fixed for a well-defined context. For instance, if the customer has an issue with financial account, then the bot has to provide follow up questions to comprehend if the account is for saving or trading, it is setup for day trading or regular trading, and so on. Therefore, the bot has to reflect the specific domain it is employing by its personality, and it shouldn't place users in a situation where they need to guess the correct answer.

COGNITIVE LOAD: Chatbots can reduce cognitive load and provide emotional link through the conversation. To reduce the cognitive load for users approaches, such as images, buttons, or structured inputs could be used. The bot have to provide clear choices, such as "Black or White" to provide clear indication on how to input information. To achieve a cognitive load in a bot, custom soft keyboards could permit a limited range of inputs and can save a bunch of typing. For example, rather than asking users to type a "Yes" or "No" text, the bot could present two mutually exclusive buttons. Moreover, introducing simplicity and avoiding complicated branching paths could decrease cognitive load.

LIMITED SCOPE: Chatbots have to focus on a well-defined problem, which means including simplification which reduces user confusion. Narrowing bot focus helps to deliver value and answer critical design questions. To limit the scope, designers should focus on task specification, predict and personalise, keep conversation short, work on triggers and actions, and avoid NLP complications.

INTERACTION TYPE: Building natural interaction is essential for bot success. While interacting with users, some processes might take time, others could be executed at once. It is a good practice to wait a bit instead of sending an immediate response. In this way, the UX will work more natural in the bot. Bots must be always ready and show some interactions similar to human beings. For instance, by displaying some emoji reactions, or a message indicating the bot is typing a text, similar to real human-human conversation. This creates a more entertaining conversation with the bot. Using short replies helps in creating positive UX by letting the user know about their query state. To illustrate, providing user with message indicating their state, such as "Done, An error occurred, I'm on it". These brief messages keep users informed about the process and the results. In addition, if the bot requires more time to complete a task, it should communicate it to the user and not keep silent.

PERSONALITY: Bots personality is part of the interface, since it decides if the bot is friendly in terms of interaction, usability, usefulness, to mention a few. Therefore, CUI designers should provide the bot with ways so that users feel a natural interaction. For example, the bot should have an escape route, validate input and allow reaching a human when necessary, especially in critical situations (e.g., healthcare). Integrating elements, such as providing random responds, conversation flow, predict and personalise, emotional and empathy states, flexibility are necessary when working on bot personality.

Table II: Domain Specific CUI Design Patterns

| Domains | Pattern Features | Elements/Techniques | Platform Example |
|---|---|---|---|
| E-commerce | Goal Fulfilment, Follow Up Questions, On-boarding, Cognitive Load, Conversation Interface | Bots Flexibility, Use Reactions for Interaction, Tasks and Duty Specification, Chatbots Personality, Text vs Custom Keyboard, Perceived usefulness, User age/gender, User Boredom, Conversation flow, Emotional State, Providing Random Responds, Simplicity in Interaction, Keep Conversation Short, Fully/Partially Automated, Security and privacy, User age/gender, User Boredom, predict and personalise, Provide A Way Out, Analytics, Curating, Convenient Bot | Amazon Echo, Operator, Allo from Google. [7] |
| Health & Lifestyle | Goal Fulfilment, Input Validation, On-boarding, Conversation Flow, Follow Up Questions, Personality Style | Bots Flexibility, Use Reactions for Interaction, Tasks and Duty Specification, Text vs Custom Keyboard, Rigid Syntax, Conversation flow, Emotional State, Providing Random Responds, Keep Conversation Short, Simplicity in Interaction, Chatbots Personality, Perceived usefulness, predict and personalise, Provide A Way Out, Analytics, Curating, Convenient Bot | ELIZA, Sense.ly, Your.MD, Florence. [8] |
| Productivity | Conversation Flow, Content, Commands, Input Validation, Conversation Interface, Cognitive Load | Simplicity in Interaction, Conversation flow, Bots Flexibility, Keep Conversation Short, Use Reactions for Interaction, Chatbots Personality, Text vs Custom Keyboard, Providing Random Responds, Rigid Syntax then NLP, Security and privacy, User Boredom, Tasks and Duty Specification, Fully/Partially Automated, NLP Complication, Simplicity in Interaction, Security and privacy, Emotional State, Analytics, Curating, Convenient Bot | API.AI. [9] |
| Finance | Cognitive Load, Follow Up Questions, Commands, Content | Text vs Custom Keyboard, Simplicity in Interaction, Emotional State, Keep Conversation Short, Empathy in Conversation Interfaces, NLP Complication, User Boredom, Chatbots Personality, Tasks and Duty Specification, Perceived usefulness, User age/gender, User Boredom, Conversation flow, Bots Flexibility, Fully/Partially Automated, Providing Random Responds, Rigid Syntax then NLP, Security and privacy, Predict and Personalise, Curating, Convenient Bot | Kasisto. [10] |
| Entertainment | Conversation Flow, Cognitive Load, Interaction Type, Limit Scope | Simplicity in Interaction, Conversation flow, Bots Flexibility, Keep Conversation Short, Use Reactions for Interaction, Text vs Custom Keyboard, Empathy in Conversation Interfaces, Emotional State, NLP Complication, User Boredom, Chatbots Personality, Notification for User Engagement, Providing Random Responds, | Telegram Bot. [11] |

## V. Conclusion

The next generation CUI can predict from our activity patterns when our meeting ends, calculate that we need a taxi, even send sensitive invitations for friends. Conversation involves little layout and design, the UX is primarily based on a sequence of messages. This paper attempted to outline important research challenges to be addressed when designing CUI for specific domains. Based on our investigation, there is a certain pattern that works better in certain domains and within a specific context. There is no one-size-fits-all approach when designing CUIs that can be applied in all domains. This paper will contribute to the existing state in CUI research by enabling designers and developers to efficiently process, manipulate and absorb best practices in CUI design. While still much work is needed on the behavioural and technical parts of conversational interface design, we hope this study to be a useful guideline to follow when designing chatbot interfaces for particular domains.